\documentclass{sig-alternate-2013}

\usepackage{amsmath}
\usepackage{graphicx}

\newfont{\mycrnotice}{ptmr8t at 7pt}
\newfont{\myconfname}{ptmri8t at 7pt}
\let\crnotice\mycrnotice%
\let\confname\myconfname%

\permission{(draft)}
\conferenceinfo{SIGCSE'15,}{March 4--7, 2015, Kansas City, MO, USA.}
\copyrightetc{~}

\newcommand{\ie}{{\em i.e.}}

\newcommand{\etal}{{\em et al.}}
\newcommand{\comment}[1]{}
\newcommand{\code}[1]{{\texttt{\small{#1}}}}

\newcommand{\term}[1]{{\ensuremath{\langle\textsf{\small{#1}}\rangle}}}

\newcommand{\keyw}[1]{\code{#1} \/}
\newcommand{\nterm}[1]{{\ensuremath{
                                    \textrm{\small{\emph{#1}}}}}}

\newcommand{\gtone}[1]{\ensuremath{{#1}^{+}}}
\newcommand{\opt}[1]{\ensuremath{{#1}_{\textit{opt}}}}

\newcommand{\alt}{\ensuremath{~\vert~}}

\begin{document}


\title{The Interpreter In An Undergraduate Compilers Course}

\numberofauthors{1}

\author{
\alignauthor
John H. E. Lasseter\\
       \affaddr{Hobart \& William Smith Colleges}\\
       \affaddr{Geneva, NY USA}\\
       \email{lasseter@hws.edu}
}

\maketitle
\begin{abstract}
An undergraduate compilers course poses significant challenges to students, in both the conceptual richness of the major components and in the programming effort necessary to implement them.  In this paper, I argue that a related architecture, the {\em interpreter}, serves as an effective conceptual framework in which to teach some of the later stages of the compiler pipeline.  This framework can serve both to unify some of the major concepts that are taught in a typical undergraduate course and to structure the implementation of a semester-long compiler project.  
\end{abstract}




\section{Introduction}

\begin{quote}
``Interpreters are the second-class citizens of the compiler construction world:  everybody 
employs them, but hardly any author pays serious attention to them.''---Grune \etal{}, {\em 
Modern Compiler Design}, p. 49
\end{quote}

The undergraduate compiler construction course at our school is a one-semester, upper-division 
elective, offered every other year.  The class is a mix of software engineering concerns and 
conceptual material from a traditional core (lexing, parsing, semantic analysis, and code 
generation), and at its heart is the construction by each student of a working compiler, 
completed over the 15-week semester.  For its balance of simplicity and realistic features, we 
focus on the Tiger programming language, originally developed by Andrew Appel for his 
project-based ``Modern Compiler Implementation'' suite of C, ML, and Java textbooks 
\cite{epita-tiger-project,appel97compilers}.  
Our projects differ from his work in that we do not tackle some of the
more advanced back-end problems such as register allocation and liveness analysis, 
concentrating instead on the simpler task of code generation for a virtual machine.  In order 
to facilitate modularity between the assignments (and prevent early difficulties from having a 
cumulative effect on later work), the students build their work from a skeleton Java source 
code distribution, consisting of some utility code, a set of AST classes, and interfaces for 
traversal of the AST.  Following the lead of more recent texts 
(\cite{appel02compilers,fischer10compilers,parr10langpatterns}, for example), this traversal 
is done according to the well-known {\em visitor pattern} \cite{gamma95designpatterns}, in 
both the semantic analysis and code generation phases.

Through several versions of this course, an unfortunate trend in student performance has 
emerged.  Students do well with the material on lexing and parsing, 
but too many of them begin to struggle during implementation of the semantic analysis phase.  
They find overwhelming both the conceptual material underlying type checking, the details of 
the visitor pattern in traversing an AST, and the interplay of this traversal technique with 
the complexities of a realistic semantic analysis implementation.  It is not uncommon for 
students to fall behind badly at this point, with the result that their final effort---AST-
based code generation---is less successful than it should be, even with the relative 
modularity of the accompanying assignments.  The reappearance of AST traversal in the code 
generation phase does serve to deepen the understanding of the semantic analysis phase, but 
often this benefit  comes only at or near the end of the semester.  

In meetings with individual students, I have often used the basic structure of a language's {\em interpreter} as a touchstone for explanation of these two phases.  Though it was not a scheduled topic in the class, this approach was one that proved almost uniformly effective among the students with whom I discussed it, so much so that I began to reorganize the classroom lectures introducing both compiler phases to include this structure as a reference point.

This paper proposes to teach this connection explicitly, through a version of the undergraduate compilers course that treats the interpreter as a first class citizen.   In this approach, the semester-long compiler construction project includes an interpreter for the source language,  which is organized in a way that makes plain the congruence between this interpreter's basic structure and that of the semantic analyzer and the code generator.   With some changes to the semester's topic schedule, a module on interpreters and their implementation is presented early in the semester, after construction of the parser, as the first substantial example of the AST traversal technique.  After a few days' study of the structure of this interpreter, the more traditional type-checking material is introduced, followed by intermediate-form translation.   Despite the apparent impracticality and distraction from ``core'' material, student learning of this topic can offer several practical and conceptual benefits.  

The key point here is not the observation that all three tasks involve a similar recursive descent of the AST.  Rather, it is an explicit emphasis on teaching both compiler stages as forms of interpretation.  This approach has the immediate practical advantage of providing a tool---the interpreter---which gives a lightweight implementation of the language's reference semantics for later testing of the completed compiler.  Moreover, it establishes an early code base that supports both reuse and the comprehension of new design and algorithmic patterns in the subsequent semantic analysis and code generation implementations.  Most importantly, it provides a lucid, unified framework in which key ideas in both type checking and code generation can be explained as forms of explicit-control evaluation over different forms of abstract value domain.  

In the remainder of this paper, we develop a model curriculum for this approach.  Section \ref{section:related} positions this work within the relevant literature.  The development proper begins in Section \ref{section:distribution} with an overview of the standard distribution for the course's semester-long project.  Section \ref{section:interpreter} introduces the primary innovation of this paper, in its presentation of an {\em interpreter} for the semester project's target language, along with a discussion of the interpreter's place in the semester schedule.   Sections \ref{section:semant} and \ref{section:codegen} discuss the semantic analysis and code generation phases, the teaching of which is the principal motivation for the development {\em a priori} of an interpreter.   There are frequent references throughout to the source code of a model semester-long project.   This code is freely available by request from the author.

\section{Related Work}
\label{section:related}

From a theoretical perspective, it is well known that both semantic analysis and code generation are closely related to a language's {\em interpreter}, the former a simulation of program execution on abstract value domains \cite{cousot97popl} and the latter a specialization of the interpreter with respect to a program's source code \cite{jones93partialeval}.  It is therefore unsurprising that the implementation of an interpreter shares many structural characteristics with the implementation of both the semantic analysis and code generation phases. However, the details of these correspondences are likely inaccessible to the typical undergraduate student, and moreover, they are a distraction.  An undergraduate course in compiler construction generally focuses on a narrow range of language features, emphasizing instead many real-world concerns such as efficient symbol table construction, separation of front and back ends through an intermediate representation, call stack frames, and (time permitting) various code-improving transformation techniques.   As a consequence, interpreters, if they are included at all, generally serve as a foil for the  superior performance of compiled code \cite{mak09compilers,watt00compilers}, or else as material for a more breadth-based course on general language implementation \cite{parr10langpatterns,scott09proglangpragmatics}. 

On the other hand, the construction of interpreters plays a prominent role in many courses on programming language design.  The simplicity of an interpreter's core structure and the close correspondence to a language's semantics makes this a natural teaching tool, both for conceptual organization and for prototypical implementation of various language features \cite{friedman08eopl,kamin90proglangs}.  Abelson and Sussman's classic CS1 text \cite{abelson96sicp} even uses this structure to introduce the structure of a simple compiler, though the correspondence between the two is quickly buried in the details of code generation, and their compiler lacks many real-world features such as a semantic analysis phase.

Several undergraduate-level texts on programming language theory make the formal connection between a language's type system and its concrete semantics explicit (for example, Harper \cite{harper13pfpl}).  The correspondence between an interpreter and a type-checker is an easy consequence of this.  

Pagan proposes the inclusion of material on partial evaluation to derive a code generator from a language's interpreter \cite{pagan89sigcse}.  However, his work in that paper is more focused on the specialization of an interpreter for a program with respect to a file of known input values, and the way in which this can be used to generate a more efficient intermediate representation (Pascal source code).  He does not address the correspondence between interpretation and semantic analysis.

\section{A Semester Compiler Project}
\label{section:distribution}

Tiger was introduced by Andrew Appel as a simple yet realistic teaching language for compiler construction \cite{appel97compilers}.  It is an imperative language, with a feature set drawn from a simplification of Pascal.  It offers a standard assortment of imperative control flow constructs, procedure definition and call, nested variable declaration scopes, and nested procedure definitions, and there is a rudimentary standard library to support string-related operations and user interaction.  Data types are limited to integers, strings, arrays, and programmer-defined records.  There is no support for either object-oriented or functional programming features.  

\begin{figure}
\[
\begin{array}{rcll}
e & := & \term{int} \alt \term{str}  \alt \keyw{nil} \alt \nterm{var}  
               & \textit{(constants, vars)}\\
  &    & \alt \nterm{var} \keyw{~:=~} e & \textit{(assignment)}\\
  &    & \alt \keyw{(}~\opt{\nterm{exp-seq}}~\keyw{)} &  \textit{(sequence)}\\
  &    & \alt e ~\nterm{op}~ e \alt \keyw{-}e & \textit{(operators)}\\
  &    & \alt \nterm{id}~\keyw{(}~\opt{\nterm{exp-list}}~\keyw{)} 
               & \textit{(procedure calls)}\\
  &    & \alt \nterm{id} ~\keyw{\{} ~ \opt{\nterm{field-list}} ~ \keyw{\}}
               & \textit{(record allocation)}\\
  &    & \alt \nterm{id} ~\keyw{\lbrack} ~ e ~ \keyw{\rbrack}~\keyw{of} ~ e
               & \textit{(array allocation)}\\
  &    & \alt \keyw{if}~e~\keyw{then}~e \\
  &    & \alt  \keyw{if}~e~\keyw{then}~e~\keyw{else}~e \\ 
  &    & \alt \keyw{while}~e~\keyw{do}~e \\ 
  &    & \alt \keyw{for}~\nterm{id}~\keyw{:=}~e~\keyw{to}~e~\keyw{do}~e \\
  &    & \alt \keyw{break} \\
  &    & \alt \keyw{let}~\gtone{\nterm{decl}}~\keyw{in}~\nterm{exp-seq}~\keyw{end}
                & \textit{(declarations)}\\
\\
\nterm{decl}
  & := & \keyw{type}~\nterm{id}~\keyw{=}~\nterm{type}\\
  &    & \multicolumn{2}{l}{
         \alt \keyw{var}~\nterm{id}~\keyw{:=}~e 
         \alt \keyw{var}~\nterm{id}\keyw{:}\nterm{id}~\keyw{:=}~e 
         } \\
  &    & \multicolumn{2}{l}{
         \alt \keyw{function}~\nterm{id}~\keyw{(}~\opt{\nterm{exp-list}}~\keyw{)}~\keyw{=}~e
         } \\
  &    & \multicolumn{2}{l}{
         \alt \keyw{function}
                 ~\nterm{id}~\keyw{(}~\opt{\nterm{exp-list}}~\keyw{) : }\nterm{id} ~\keyw{=}~e
          } \\
\\
\nterm{type}
  & := & \nterm{id}  & \textit{(type aliases)}\\
  &    & \alt \keyw{\{} ~\nterm{type field list} ~\keyw{\}} & \textit{(record types)}\\
  &    & \alt \keyw{array of~} \nterm{id} & \textit{(array types)}\\
\end{array}
\]
\caption{Tiger language syntax (abbreviated)}
\label{fig:tiger-syn}
\end{figure}

Following Appel's lead, students in our course implement a compiler for Tiger from a skeleton distribution, consisting of some utility classes, a collection of classes representing the possible types, and interfaces for important components of code generation, such as stack frames, abstractions of frame access, labels, and so on.  There is a complete set of AST classes, corresponding to the concrete grammar illustrated in Figure \ref{fig:tiger-syn}.  

Finally, the distribution includes a pair of interfaces to support traversal of according to the {\em visitor pattern}.  Behaviors that are defined by traversal of an AST implement the appropriate {\em visit()} methods:

{\footnotesize
\begin{verbatim}
public interface IAbsynVisitor {
    void visit(ExpInt e);
    void visit(ExpIfElse e);
    void visit(ExpLet e);
    void visit(ExpWhile e);

          ... (etc.)
}
\end{verbatim}
}

\noindent  In turn, each of the AST node classes includes an {\em accept()} method

{\footnotesize
\begin{verbatim}
public interface IVisitable {
    void accept(IAbsynVisitor v);
}
\end{verbatim}
}
\noindent with each {\em accept()} implementation dispatching the appropriate behavior ({\ie}, {\em visit()} call) for an object's type.  For example, 

{\footnotesize
\begin{verbatim}
public class ExpLet implements IVisitable {
          ... 
    public void accept(IAbsynVisitor v) {
        v.visit(this);
    }
          ...
}
\end{verbatim}
}

A common motivation behind the use of this pattern is the need to defer the definition of one or more operations on a collection of data without having to later modify the definitions of those classes themselves, and at the same time providing some compile-time assurance that the behavior definition is exhaustive on the collection of class definitions.  This need to separate future behavioral commitments from a fixed set of data types is particularly acute in the case of a student compiler project, as students must learn the AST hierarchy weeks before they encounter the last stages of the compiler.  

The disadvantage is any implementation of a visitor pattern is, at best, complicated, and the first encounter with this pattern and its multiple redirection of control often proves confusing.  To ameliorate this confusion, students first encounter AST traversal through the visitor pattern as they finish learning basic parsing and syntax-directed translation.  At this point, they have each built a working grammar specification file with production rules whose associated actions result in the construction of an AST from the concrete source code.  Among the tools in the standard distribution is a fully implemented pretty-print visitor, which is used to test the generated parsers.  The final day in which we study syntax-directed translation is actually dedicated to the study of this pretty-printer, and through this, students are given an introduction to the visitor pattern.

\section{Interpreter Components}
\label{section:interpreter}

Once parsing is finished, it seems to be customary to move immediately into the next stage of the compiler pipeline, semantic analysis.  This is the approach given in project-based texts  such as Fischer/LeBlanc/Cytron \cite{fischer10compilers} and Appel \cite{appel02compilers}, for example, and it is the one we have until recently used in our course.

At this point in the class, however, it is possible to add a few days of material on interpretation and the structure of an interpreter's implementation.  While challenging in its own way, a basic interpreter is considerably simpler to understand than a compiler for the same language, particularly if real-world concerns of performance are ignored.  Aside from the AST itself, the critical components of interpretation are a domain of runtime values produced by programs, an understanding of the traversal algorithm underlying interpretation, and an {\em environment} \ie, a record of the bindings of current names to their values.  

The environment is necessary to support the declaration and subsequent use of identifiers, whether they refer to variables, procedures, or type aliases.  In a compiler, this is the {\em symbol table}, and efficient construction of these can involve some complicated data structures.  In its essence, however, it has a very simple interface, needing only the ability to retrieve the value associated with an identifier, add new identifier/value bindings, and record the entry to and exit from a scope:

{\footnotesize
\begin{verbatim}
public class Table<T> {
    public T get(Symbol key) { ... }
    public void put(Symbol key, T value) { ... }

    public void beginScope() { ... }
    public void endScope() { ... }
}
\end{verbatim}
}

Our distribution includes a ``semant'' package that provides students with a complete implementation of this class, along with a ``wrapper'' class, {\tt Env}, which adds on a few utility methods.  

{\footnotesize
\begin{verbatim}
public abstract class Env<T> {
    public final Table<T> env;

    protected static Symbol sym(String s) {
        return Symbol.symbol(s);
    }
       ... (etc.)
}
\end{verbatim}
}

Typically, a compilers text will cover the problem of symbol table data structures in some detail, though many courses (including ours) elide this.  The project distribution used in our class realizes a fairly efficient implementation using hashtables and ``binding'' objects that support fast addition and disposal of scopes.  However, if one wanted students to gain experience implementing such tables on their own, a more straightforward, stack-based prototype could be provided at this point instead.

The declaration of the {\tt Env} class as both generic and \linebreak
{\tt abstract} reflects the broader theme of this project:  each of the ``inter\-preter-like'' stages of the compiler, semantic analysis, and code generation, will have its own version of an environment.  Yet where the interpreter uses the environment to store symbol/value bindings, the abstract forms of interpretation will store bindings of symbols to {\em value abstractions}.  In our work, this is reflected in the fact that the interpreter, semantic analyzer, and code generator all define a comparable hierarchy of {\tt Entry} classes, reflecting the differences across the three uses of the environment.  Following Appel \cite{appel97compilers}, we use an abstract {\tt Entry} superclass and two child entries, {\tt VarEntry} and {\tt FunEntry}:

\begin{center}
\includegraphics[scale=0.5]{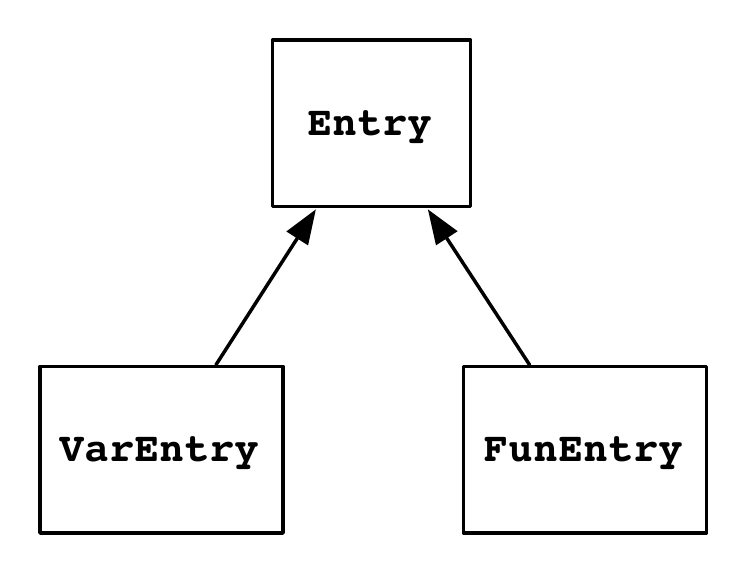}
\end{center}

\noindent  The difference is that we have a separate version of this hierarchy for each of the interpretation, semantic analysis, and code generation components.  Although each of these stages needs to distinguish bindings of procedure names from ordinary value bindings, the difference in their respective levels of abstraction means that, for example, a {\tt VarEntry} (\ie, name/value binding) for an interpreter means something different from the corresponding {\tt VarEntry} during code generation.  

It is these {\tt Entry} families that parameterize {\tt Env} declarations, in the concrete environment designed for a stage.  For example, our interpreter uses a representation of ``value'' (for Tiger, this means integers, strings, records, or arrays) to define the notion of a {\tt VarEntry} and {\tt FunEntry}, which leads in turn to

{\footnotesize
\begin{verbatim}
public class EnvInt extends 
     tigerc.semant.Env<tigerc.semant.interp.Entry> { 
     ... 
}
\end{verbatim}
}

Finally, there is the interpreter itself.  The version developed for this paper is a single class, which includes a value environment ({\tt EnvInt}), a utility error message handler, and which implements the AST visitor ({\em IAbsynVisitor}), with the {\tt result} field storing the `return'' value of each {\tt visit()} call:

{\footnotesize
\begin{verbatim}
public class InterpV implements IAbsynVisitor {
    private IValue result = null;
    private EnvInt env; 
         ...
    public void visit(ExpInt e) {
        result = new ValInt(e.value);
    }
    public void visit(ExpOp e) {
        e.left.accept(this);
        IValue v1 = this.result;
        e.right.accept(this);
        IValue v2 = this.result;

        switch (e.oper) {
        case PLUS: { 
          this.result = 
              new ValInt(((ValInt) v1).val 
                          + ((ValInt) v2).val);
          break;
        }
          ...
        }
    }
    public void visit(ExpLet e) {
        env.env.beginScope();
        for (Decl dec : e.decls) {
            dec.accept(this);
        }
        e.body.accept(this);
        env.env.endScope();
    }

    ... (etc.)
}
\end{verbatim}
}

It is worth noting the problem of type safety here.  Although Tiger is a statically typed language, students have not yet learned anything about semantic analysis.  Two options are possible here.  On the one hand, we can simply ``cheat'' on the language specification and add easy dynamic checks of safety to the {\tt visit()} bodies (or, worse, we can ignore the problem altogether).  Alternately, we can use the problem of unsafe methods in the interpreter to motivate type checking, with students diving in immediately to the implementation of semantic analysis, though applied here to the interpreter.  At present, it is unclear, which of these leads to a smoother outcome in learning semantic analysis implementation.

The {\tt InterpV} class can be given to students as a complete product, with a couple of days devoted to studying its details.  Alternately, one might choose a partial distribution, with the expectation that they complete it on their own.  However, the purpose of this digression in the course is to provide a richer basis of study and comprehension for the actual later compiler stages, so it is important that students finish this component of the course with a fully functional interpreter to study, whether that is provided {\em a priori} or {\em a posteriori}.

\section{Semantic Analysis}
\label{section:semant}

Semantic analysis is the last ``filter'' of bad programs before the back end phase(s), and languages of a realistic scale (even one as small as Tiger) include a complicated array of requirements:  compatibility of value types with operators, declaration of variables before use (and use only within a given scope), non-assignability of bounded loop counters, enclosing loops for nonstandard jumps such as {\tt break}, non-circularity of type alias definitions, and so on.  At root, however, all of these verification tasks can be understood as the simulation of program execution on a suitable set of abstractions of the possible concrete values.  This is not only true as a theoretical statement \cite{cousot97popl}, it is a useful lens for comprehension of the type checking process.

Our implementation of semantic analysis is structured to reflect this view in a simple manner.  \`{A} la Appel, the semantic analysis package ({\tt tigerc.semant.analysis} ) makes use of a collection of ``type'' objects, ({\tt tigerc.semant.types}), instead of the {\tt IValue} hierarchy of the interpreter.  Similarly, it includes the following {\tt Entry} hierarchy:

{\footnotesize
\begin{verbatim}
public interface Entry {}

public class VarEntry implements Entry {
    public final Type ty;
    public final boolean assignable;
       ...
}

public class FunEntry implements Entry {
    public final List<Pair<Symbol, Type>> formals;
    public final Type result;
       ...
}
\end{verbatim}
}

\noindent These value abstractions are used to implement an environment of symbol/type bindings:

{\footnotesize
\begin{verbatim}
public class EnvTC extends 
  tigerc.semant.Env<tigerc.semant.analysis.Entry> {

    protected final Table<Type> tenv;  
    public EnvTC() {
        tenv = new Table<Type>();

        // Bindings for the primitive types.

        tenv.put(sym("int"), PrimTy.INT_T);
        tenv.put(sym("string"), PrimTy.STRING_T);

        makeStdLib();
    }
         ... (etc.)
}
\end{verbatim}
}

\noindent  This version of the environment specializes its generic parent with its own, abstract notion of {\tt Entry}.  Hidden inside is a second symbol table, to support Tiger's requirement that an identifier can exist as both a variable and a type name 
definition. 

With these abstractions of value (the {\tt semant.types} classes) and environment, we are ready to implement semantic analysis in a form highly congruent to that of the interpreter:

{\footnotesize
\begin{verbatim}
public class SemantV implements IAbsynVisitor {

    private Type ty = null;  // the "return value"
    private EnvTC env;
    private ErrorMsg err;
         ...
    public void visit(ExpInt e) {
        this.ty = new INT();
    }
    
    public void visit(VarSimple v) {
        Entry en = env.env.get(v.name);
        if (en == null) {
            err.error( ... );
            this.ty = ERROR.instance();
        } else if (en instanceof FunEntry) {
            err.error( ... );
            this.ty = ERROR.instance();
        } else {
            VarEntry var = (VarEntry) en;
            this.ty = var.ty; 
        }
    }

    ... (etc.)
}
\end{verbatim}
}

\section{Code Generation}
\label{section:codegen}

In its industrial-strength form, code generation is the most formidable of the stages in a compiler.  When a physical architecture is targeted, it is common to translate the AST into an intermediate representation that, although still tree-structured, more closely resembles the assembly language of real machines.  Even in the case where the target language is a virtual machine like the JVM, a realistic compiler must perform extensive data flow analysis and code-improving transformations.

At the core, though, is a partial evaluation of the interpreter:  literally, a traversal of the AST, for which the result is the machine code that corresponds to the execution steps the interpreter would take.  This is true whether our target is virtual machine code or an intermediate tree language.  

In this stage, the abstraction of value is not a type but rather a description of the resources that must be laid out for a value, either a register or a stack frame.  For our project, that means an abstraction of {\em frames} and {\em access} within a frame:

{\footnotesize
\begin{verbatim}
public interface IFrame {
    IAccess allocLocal(Type t);

    IAccess popLocal();
    // removes and returns the resource allocated at 
    // the end of this frame.

    int frameEnd();
    //offset of the first available word in this frame
}

public interface IAccess {
    int offset();
    Type getType();
}

\end{verbatim}
}

To aid in calculating access properties on some machines, we will need to make use of types here, though we only need record them, rather than checking, since they are all presumed correct at this point.  This gives rise to an environment definition abstracted as

{\footnotesize
\begin{verbatim}
public class EnvTrans extends 
               tigerc.semant.Env<Entry> {
    public final Table<Type> tenv;
     
    ... (etc.)
}

public class VarEntry implements Entry {
    public final Type ty;
    public final IAccess access; 
     ...
}

public class FunEntry implements Entry {

    public final List<Pair<Symbol, Type>> formals;
    public final Type result;

    private Label label;
    private Class<?> extern;
    private IFrame frame;
    ...
}
\end{verbatim}
}

Unfortunately, there are a some practical issues that arise with a straightforward analogy to interpreters here, primarily in the need to distinguish  more carefully the use of a variable as an l-value or r-ralue.  The code generated for these is often quite different.  For example, suppose we are targeting JVM code (as our project does).  For simple identifiers, we don't need to put the lefthand-side (LHS) value on the stack.  It suffices to compute the righthand side (RHS), then store the value.  For array subscripts and record fields, however, we need to do a little more work in computing the LHS.  Too, the address on the record/array must be put on the stack, just below the top (which will contain the value of the RHS). Nonetheless, the interpreter/code generation correspondence is very close:

{\footnotesize
\begin{verbatim}
public class JVMGeneratorV implements 
          tigerc.semant.translate.ICodegen,
            tigerc.syntax.absyn.IAbsynVisitor {


    private IFrame frame = new JVMFrame();
    private EnvTrans env; 
    private java.io.PrintWriter tgtOut; 
    // Target destination for generated code
    private Type expType; 
    // type of the most recently-visited expression

    public void visit(ExpInt e) {
        emitLn(this.code, "ldc " + e.value);
        this.expType = PrimTy.INT_T;
    }
    
    public void visit(ExpAssign e) {
        if (e.lhs instanceof VarSimple) {
            e.rhs.accept(this);
            
            this.rvalueMode = false;
            e.lhs.accept(this);
            this.rvalueMode = true;
        } else {
            ... 
        }
        this.expType = PrimTy.VOID_T;
    }

    public void visit(VarSimple x) {
        VarEntry v = (VarEntry) env.env.get(x.name);
        String loadstore = 
            (this.rvalueMode ? "load" : "store");

        if (v.ty.coerceTo(PrimTy.INT_T)) {
            emitLn(this.code, "i" + loadstore 
                      + " " + v.access.offset());
        } else { 
            emitLn(this.code, "a" + loadstore 
                      + " " + v.access.offset());
        }
        this.expType = v.ty.actual();
    }
        ...
}
\end{verbatim}
}

\section{Conclusion and Future Work}

This paper has shown how to use a known set of theoretical connections between a language's interpreter and key stages in the back end of a compiler for that language to structure the last few weeks of a compilers course.  It provides both an elegant, reusable body of code and a set of patterns through which students can organize their understanding of these topics.

Future opportunities include the assessment of this approach among a larger population of students, as our own institution is small enough to make substantial data gathering of this kind impractical.  Too, there remains the question of other opportunities within the pipeline in which to leverage this analogy.  In more advanced courses, for example, is it helpful to bring in some of the original insights from abstract interpretation, in order to explain the task of data flow analysis?  Is there a corresponding notion of interpretation of machine code versus abstractions of interpretation on tree-valued intermediate form?  The prospect of future investigation in these directions is exciting.


\bibliographystyle{abbrv}


\end{document}